\documentclass[aip, preprint]{revtex4-1}
\usepackage{graphicx}

\begin{document}

\title{Dynamic Interface Rearrangement in LaFeO$_3$ / $n$-SrTiO$_3$ Heterojunctions}

\author{Steven R. Spurgeon}
\affiliation{Physical and Computational Sciences Directorate, Pacific Northwest National Laboratory, Richland, Washington 99352, USA}
\email{steven.spurgeon@pnnl.gov; sa.chambers@pnnl.gov}

\author{Peter V. Sushko}
\affiliation{Physical and Computational Sciences Directorate, Pacific Northwest National Laboratory, Richland, Washington 99352, USA}

\author{Ryan B. Comes}
\affiliation{Physical and Computational Sciences Directorate, Pacific Northwest National Laboratory, Richland, Washington 99352, USA}
\affiliation{Department of Physics, Auburn University, Auburn, Alabama 36849, USA}

\author{Scott A. Chambers}
\affiliation{Physical and Computational Sciences Directorate, Pacific Northwest National Laboratory, Richland, Washington 99352, USA}

\date{\today}

\begin{abstract}

Thin film synthesis methods developed over the past decades have unlocked emergent interface properties ranging from conductivity to ferroelectricity. However, our attempts to exercise precise control over interfaces are constrained by a limited understanding of growth pathways and kinetics. Here we demonstrate that shuttered molecular beam epitaxy induces rearrangements of atomic planes at a polar / non-polar junction of LaFeO$_3$ (LFO) / $n$-SrTiO$_3$ (STO) depending on the substrate termination. Surface characterization confirms that substrates with two different (TiO$_2$ and SrO) terminations were prepared prior to LFO deposition; however, local electron energy loss spectroscopy measurements of the final heterojunctions show a predominantly LaO / TiO$_2$ interfacial junction in both cases. \textit{Ab initio} simulations suggest that the interfaces can be stabilized by trapping extra oxygen (in LaO / TiO$_2$) and forming oxygen vacancies (in FeO$_2$ / SrO), which points to different growth kinetics in each case and may explain the apparent disappearance of the FeO$_2$ / SrO interface. We conclude that judicious control of deposition timescales can be used to modify growth pathways, opening new avenues to control the structure and properties of interfacial systems.

\end{abstract}

\maketitle

The predictive design and synthesis of robust functional interfaces represents the next frontier for oxide materials. While many model systems have been explored theoretically and experimentally, it is increasingly apparent that an insufficient understanding of growth pathways poses a roadblock to achieving predicted properties in real systems. Previous studies have identified several defect factors that affect the structure of oxide interfaces, including misfit strain,\cite{SankaraRamaKrishnan2014} oxygen vacancies,\cite{Nakagawa2006} cation intermixing,\cite{Chambers2010} and the migration of entire lattice planes.\cite{Lee2014} This last mechanism has been called ``dynamic layer rearrangement`` by Lee \textit{et al.}, who found that it is energetically favorable for SrO $\leftrightarrow$ TiO$_2$ planar swapping to occur during the synthesis of ($A$O)($B$O$_3$)$_n$ Ruddlesden-Popper phases;\cite{Lee2014, Nie2014} their analysis prompted the use of a floating SrO surface layer to stabilize the incorporation of TiO$_2$ planes into the desired Sr$_2$TiO$_4$ phase. Saint-Girons \textit{et al.} also invoked a SrO $\leftrightarrow$ TiO$_2$ rearrangement mechanism to explain the coalescence of SrTiO$_3$ (STO) islands grown on Si, which they termed a ``knitting machine'' process.\cite{Saint-Girons2016} These studies show how the mobility of lattice planes during non-equilibrium growth processes can introduce significant, potentially useful, deviations from an ideal structure. A careful experimental investigation of the evolution of the near-surface region, in conjunction with growth pathway modeling, may offer unique insight into the stability and synthesis of oxide heterojunctions.

Oxide molecular beam epitaxy (MBE), with its ability to produce single-crystalline thin films a monolayer at a time,\cite{Chambers2000,Chambers2010b} represents the ideal method to explore the dynamic rearrangement process. In contrast to the co-evaporative nature of other techniques, such as pulsed laser deposition (PLD), MBE permits shuttering of elemental sources, which provides an additional degree of freedom to control stoichiometry and layer configurations. Layer-by-layer deposition of films and heterostructures offers the unique opportunity to harness kinetic controls by untangling processes that occur under conditions far from equilibrium. Here we consider the application of the shuttered growth mode to the synthesis of polar / non-polar LaFeO$_3$ (LFO) / $n$-STO interfaces, which have been shown to possess a suitable bandgap and built-in potential for photochemical water splitting.\cite{Stoerzinger2017, Comes2016a, Mark2014} M. Nakamura \textit{et al.} reported properties of LFO grown on SrO-- and TiO$_2$--terminated bulk STO substrates using PLD;\cite{Nakamura2016} the authors observed different film polarization states in the heterojunctions, which they attributed to differing local dipole magnitudes. A subsequent photoconductivity study\cite{Nakamura2016a} of PLD-grown LFO / STO heterojunctions by K. Nakamura \textit{et al.} also proposed the existence of different interface charge states. Comes and Chambers evaluated the effect of substrate termination directly.\cite{Comes2016a} To this end, they analyzed the electronic structure of LFO grown on SrO-- and TiO$_2$--terminated STO layers using shuttered MBE growth; while the authors confirmed the appropriate terminations using angle-resolved X-ray photoelectron spectroscopy (AR-XPS), they observed a negligible impact on the potential gradient across the heterojunction. These conflicting results call for further investigation into the LFO growth mechanisms that influence the properties of the heterojunction.

We use a combination of aberration-corrected scanning transmission electron microscopy (STEM) and \textit{ab initio} simulations to probe the atomic-scale configuration of the LFO / $n$-STO interface. Electron energy loss spectroscopy (STEM-EELS) measurements reveal interfaces with no planar defects that appear to have LaO / TiO$_2$ compositions in both cases, in contrast to the clear differences between TiO$_2$ and SrO terminations confirmed by AR-XPS prior to LFO deposition.\cite{Comes2016a} Fine structure measurements show no evidence for oxygen vacancies, but we do observe some Fe valence changes at the interface. To interpret these results, we conduct \textit{ab initio} simulations of the initial stages of LFO adsorption onto STO, which suggest that FeO$_2$ / SrO is highly unstable and can transform into a LaO / TiO$_2$ interface. Swapping of the surface plane during film deposition can replace the terminal STO plane, pushing the system toward a LaO / TiO$_2$-like interface configuration for both cases. This mechanism may partially erase the different potential gradients between the samples, explaining the discrepancy among previous reports of the interface electronic structure. Our results illustrate how the surface stability of alloying elements can give rise to unexpected heterojunction configurations. We advocate a rational design approach that considers the thermodynamic and kinetic factors associated with different substrate terminations to achieve specific synthesis outcomes.

\begin{figure}[h!]
\includegraphics[width=\textwidth]{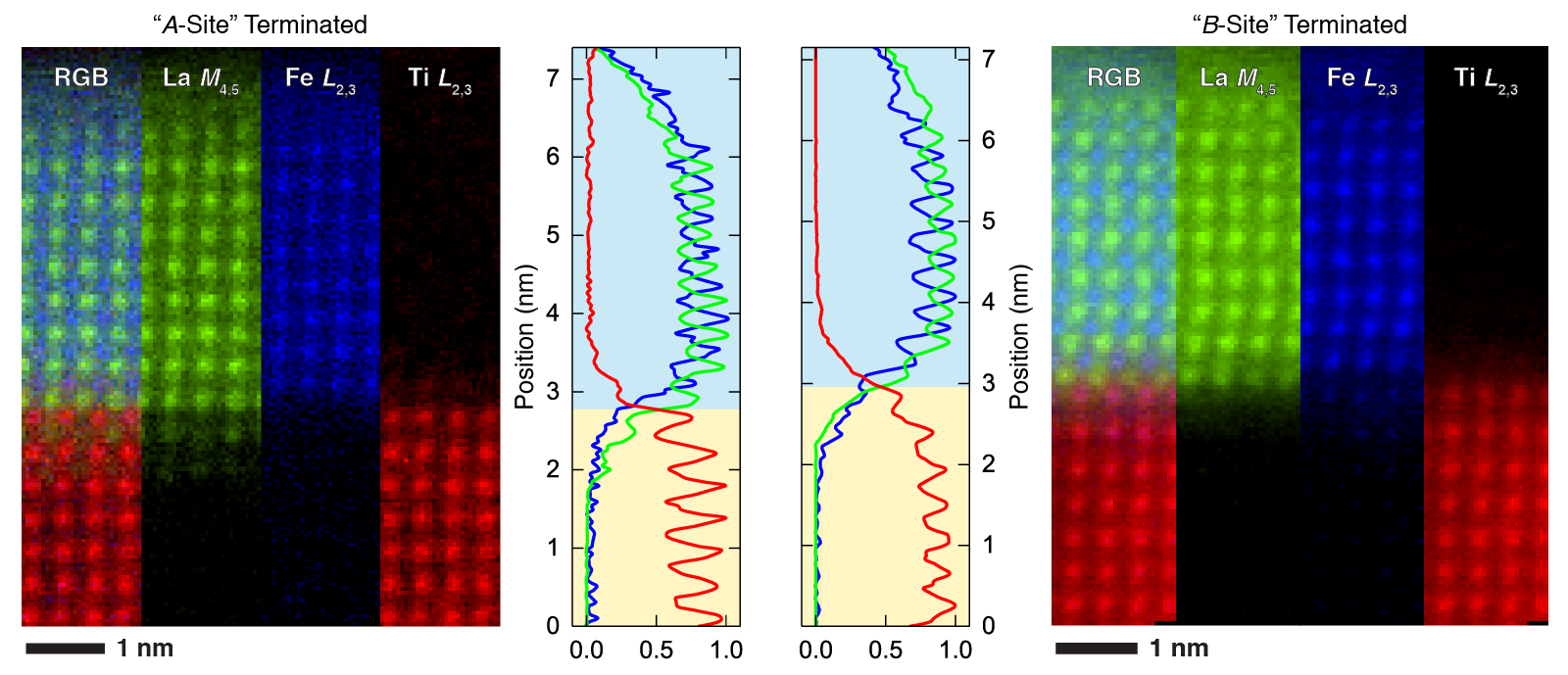}
\caption{Cross-sectional STEM-EELS mapping of local composition. PCA-filtered composite, La $M_{4,5}$, Fe $L_{2,3}$, and Ti $L_{2,3}$ integrated signal maps, alongside the integrated line profiles for the nominal ``$A$-Terminated'' (left) and ``$B$-Terminated'' (right) samples, respectively. The shaded regions mark the film-substrate interface. \label{eels_composition}}
\end{figure}

Figure \ref{eels_composition} shows representative STEM-EELS composition maps for the nominally ``$A$-Terminated'' and ``$B$-Terminated'' substrates. The most obvious feature is the similarity of the maps; in particular, both junctions have the same LaO / TiO$_2$ structure. This observation contrasts to the clear difference between the substrates prior to film deposition, as determined using AR-XPS and reflection high-energy electron diffraction (RHEED).\cite{Comes2016a} The integrated line profiles averaged in the plane of the film confirm a FeO$_2$ / LaO / TiO$_2$ / SrO interface stacking sequence for both samples, a finding consistent across multiple film regions and further confirmed in the Sr maps presented in the supporting information. The composition of the interfaces is very similar, with less than 3 u.c. intermixing for all species. This range is well within the expected delocalization of the EELS signal,\cite{Oxley2007, Spurgeon2016b} meaning that the profiles may actually be 1--2 u.c. sharper than measured. For the ``$A$-Terminated'' sample, we find that Ti drops off to a negligible level just 1 u.c. into the LFO, while it extends 2 u.c. for the ``$B$-Terminated'' sample.  Similarly, we find that the Fe signal extends just 1 u.c. into the STO for the ``$A$-Terminated'' sample, while it penetrates 3 u.c. for the ``$B$-Terminated'' sample. It is possible that Fe\textsubscript{Ti} substitution could introduce some free carriers into the system.\cite{Steinsvik1997} However, previous investigations of LaCrO$_3$ / STO superlattices have shown that similar Cr\textsubscript{Ti} substitutions have minimal effect on the electrostatic potential,\cite{Comes2017, Comes2016} a finding in line with our AR-XPS measurements of these samples.\cite{Comes2016a} Conversely, the La signal extends 3 u.c. into the STO for the ``$A$-Terminated'' sample and only 2 u.c. for the ``$B$-Terminated'' sample. La\textsubscript{Sr} substitution could also lead to electron doping of the STO, affecting the local electric field, but the low level of intermixing is unlikely to account for the dramatic change in the terminal STO layer.

\begin{figure}
\includegraphics[width=0.79\textwidth]{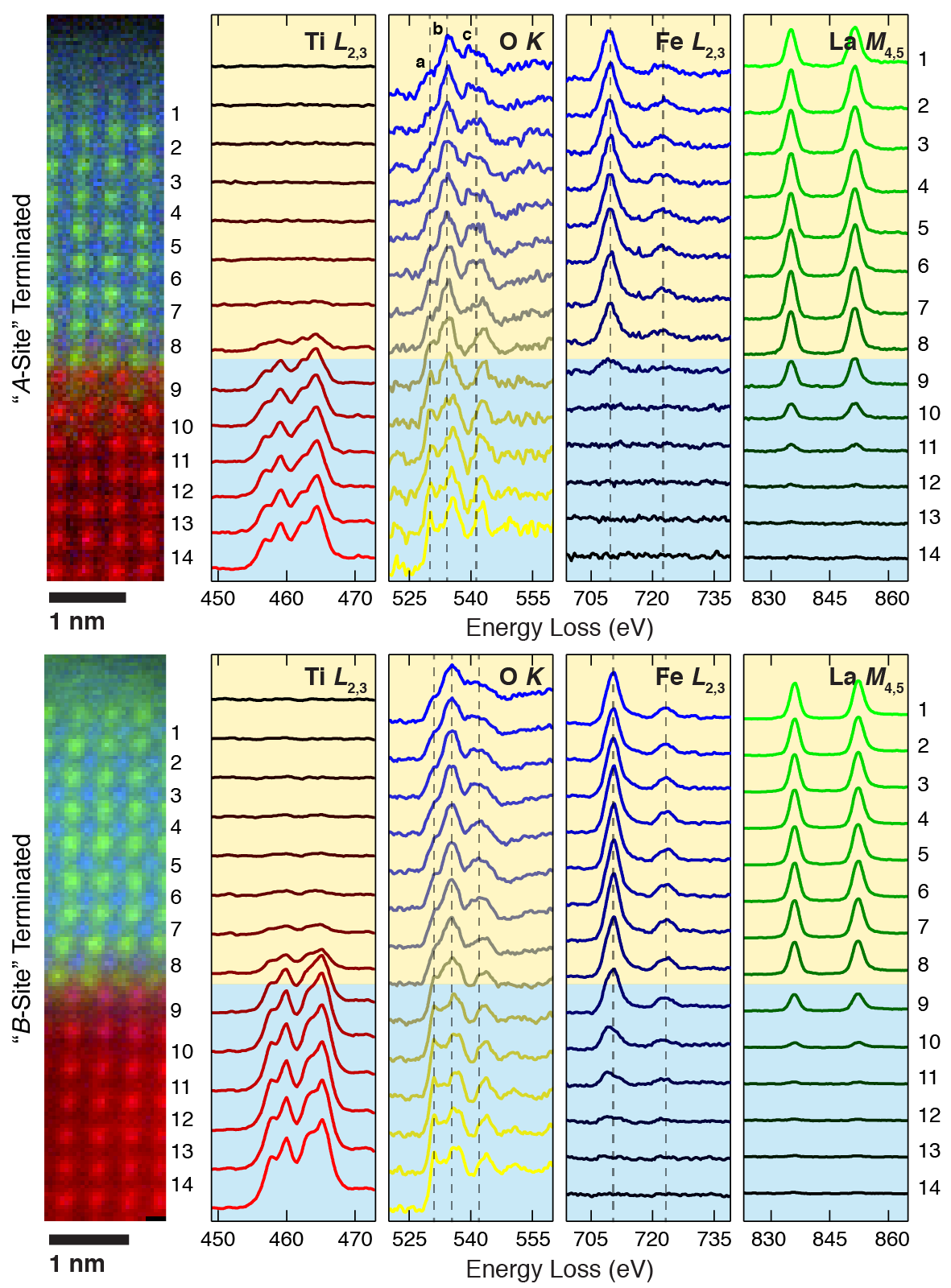}
\caption{Cross-sectional STEM-EELS mapping of local chemistry. PCA-filtered composition map and spectra for the Ti $L_{2,3}$, O $K$, Fe $L_{2,3}$, and La $M_{4,5}$ edges extracted from the unit cells labeled 1--14 for the nominal ``$A$-Terminated'' (top) and ``$B$-Terminated'' (bottom) samples, respectively. The shaded regions indicate the film-substrate interface and the dashed lines have been added as guides to the eye. The spectra have been treated to remove X-ray spikes and power-law background subtracted, but are otherwise not denoised. \label{eels_fs}}
\end{figure}

We next investigate spectral features in the EELS fine structure that would indicate chemical state changes and may point to a mechanism for interfacial reconstruction. Figure \ref{eels_fs} shows the Ti $L_{2,3}$, O $K$, Fe $L_{2,3}$, and La $M_{4,5}$ edges extracted from each u.c. across the interface for the nominally ``$A$-Terminated'' and ``$B$-Terminated'' substrates, respectively. We find that the Ti $L_{2,3}$ edge line shape is preserved from the STO bulk all the way up to the interface, within the 0.75 eV effective absolute energy resolution of our map. We observe no change in the Ti $t_{2g}$ and $e_g$ peaks that would indicate a transition from Ti$^{4+}$ to Ti$^{3+}$ valence states.\cite{Mosk2004}

The O $K$ edge is highly sensitive to local bonding in perovskite oxides,\cite{Varela2009} with the distribution of spectral features acting as a guide to local chemical states. Inspection of Figure \ref{eels_fs} reveals three distinct edge features (labeled \textbf{a}-\textbf{c}), resulting from the hybridization of O $2p$ states with $B$-site $3d$, La $5d$, and $B$-site $4sp$ bands, respectively.\cite{Varela2009} We find that the pre- to main-peak (\textbf{a}/\textbf{b}) ratio, a known indicator of valence changes, remains largely constant on the LFO side for both terminations; similarly, there is no broadening of the main peak \textbf{b} that would indicate the formation of oxygen vacancies at the interface.\cite{Nord2015} On the STO side, we do observe a small decrease in the \textbf{a}/\textbf{b} ratio moving from the bulk to within 2 u.c. of the interface for the ``$B$-Terminated'' sample, suggesting some $B$-site valence modification; similar behavior is present in the (slightly noisier) spectra for the ``$A$-Terminated'' sample. 

Turning to the Fe $L_{2,3}$ edge, we first note that the edge position remains unchanged throughout the LFO for both substrate terminations. However, we observe a measurable Fe signal 1 u.c. into the STO (spectrum 9) for the ``$A$-Terminated'' sample and a Fe signal 3 u.c. into the STO (spectra 11) for the ``$B$-Terminated'' sample. In the latter sample, we measure a 1.75 eV shift of the Fe $L_3$ edge to lower energy loss beginning at spectrum 10, which indicates a slight reduction in Fe valence toward a more Fe$^{2+}$-like state.\cite{Schmid2006} To the best of our knowledge, electron transfer to the Fe ions in this system has not been previously observed, but this behavior agrees well with changes present in the O $K$ edge spectra, as well as predictions of interface conductivity.\cite{Xu2017} Finally, although we do not detect any significant changes in the La $M_{4,5}$ edge line shape, we do find that its signal penetrates 1 u.c. deeper in the case of the ``$A$-Terminated'' sample (a finding present in multiple maps). In summary, we observe no clear modification of Ti valence, but features of the O $K$ and Fe $L_{2,3}$ edge spectra support a slight reduction in Fe valence within the top 3 STO u.c. for the ``$B$-Terminated'' sample. While some changes in the O $K$ edge are also present in the ``$A$-Terminated'' sample, we observe no comparable shift of the Fe $L_{2,3}$ edge.

Our STEM-EELS composition maps reveal minimal intermixing with no apparent long-range diffusion that would give rise to the rearrangement of the terminal STO layer. We find no evidence for oxygen vacancies and observe only a slight Fe valence reduction, which our prior AR-XPS results show to have negligible impact on the measured potential.\cite{Comes2016a} The reduction in Fe valence without oxygen vacancies is to be expected for Fe$^{3+}$ at the interface or alloyed with $n$-doped STO due to the propensity of Ti$^{3+}$ ions to transfer charge to Fe$^{3+}$, which has been observed in LaTiO$_3$ / LaFeO$_3$ interfaces\cite{Kleibeuker2014} and Sr$_{1-2x}$La$_{2x}$Ti$_{1-x}$Fe$_x$O$_3$ films.\cite{Comes2016b} This transfer leads to the formation of Fe$^{2+}$, which is observed in the EELS data and in the previous works.

\begin{figure}[h!]
\includegraphics[width=\textwidth]{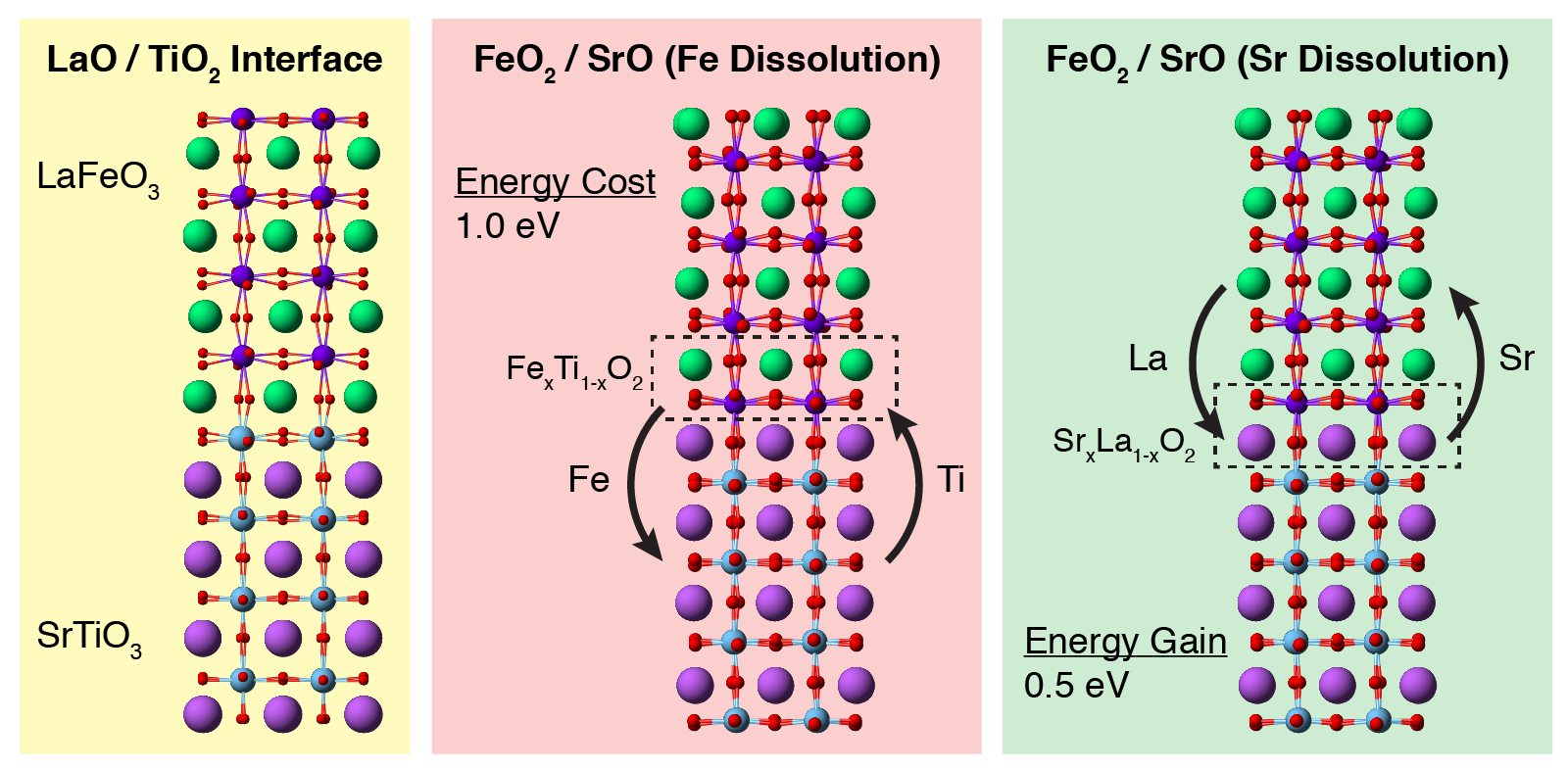}
\caption{Possible interface rearrangements and corresponding calculated energies. The left panel shows the stable case of the LaO / TiO$_2$ interface, while the middle and right panels show energetically unfavorable and favorable FeO$_2$ / SrO interface configurations, respectively. The listed energies are calculated for the $x$ = 0.25 composition. Green = La, magenta = Sr, blue = Fe, cyan = Ti, and red = O atoms.\label{dft}}
\end{figure}

To rationalize the disappearance of the FeO$_2$ / SrO interface, we examined the stability of LFO grown on SrO-- and TiO$_2$--terminated STO using {\it ab initio} simulations, as shown in Figure \ref{dft}. The relative stability of the FeO$_2$ / SrO and LaO / TiO$_2$ interfaces was evaluated in terms of separation energies of the LFO and STO slabs. We found that FeO$_2$ / SrO interface is 0.86 eV per 1$\times$1 lateral cell is less stable than LaO / TiO$_2$. A similar energy difference (0.8 eV per cell) was obtained by calculating the energy gain from depositing 1 u.c. of LFO on 5 u.c. thick TiO$_2$--terminated \textit{versus} on 5.5 u.c. thick SrO--terminated slabs of STO. The lower stability of the FeO$_2$ / SrO interface is consistent with its apparent absence in the case of LFO grown on SrO--terminated STO (c.f. Figure \ref{eels_composition}).

Examination of the middle and right panels of Figure \ref{dft} shows that FeO$_2$ / SrO can transform into a LaO / TiO$_2$ interface if either Fe or Sr at this interface is replaced with Ti or La, respectively. To assess the likelihood of such transformations, we considered several configurations that correspond to Ti---Fe intermixing so that between 50 and 75 \% of the Fe species at the FeO$_2$ / SrO interface are dissolved into STO. These configurations include Fe atoms arranged in the column and screw with 25\% site occupancy and in two neighboring and next-neighboring planes with 50\% site occupancy. Equivalent configurations were considered for Sr dissolved over La sites in the LFO part of the slab. Our calculations suggest that, in the case of Sr, such intermixing results in an energy gain of up to 0.5 eV per 2$\times$2 cell, while equivalent Fe---Ti intermixing configurations correspond to an energy cost over 1 eV. This trend is consistent with previous reports on Ti---Al intermixing at the LaO / TiO$_2$ interface in LaAlO$_3$ / STO,\cite{Chambers2010} where the opposite interface polarity mismatch is compensated by intermixing of the $B$-site species.

While intermixing is present for both FeO$_2$ / SrO and LaO / TiO$_2$ interfaces, its effect on interface structure is more pronounced in the former. Here we consider the initial stages of the LFO growth to understand the origin of this asymmetry. Since MBE deposition of LFO was performed in the shuttered growth mode, we assume that the asymmetry in the structure of FeO$_2$ / SrO and LaO / TiO$_2$ interfaces is mainly determined by sequential deposition of the first Fe- (in the case of FeO$_2$ / SrO) and La-containing (in the case of LaO / TiO$_2$) planes, respectively.

\begin{figure}[h!]
\includegraphics[width=0.8\textwidth]{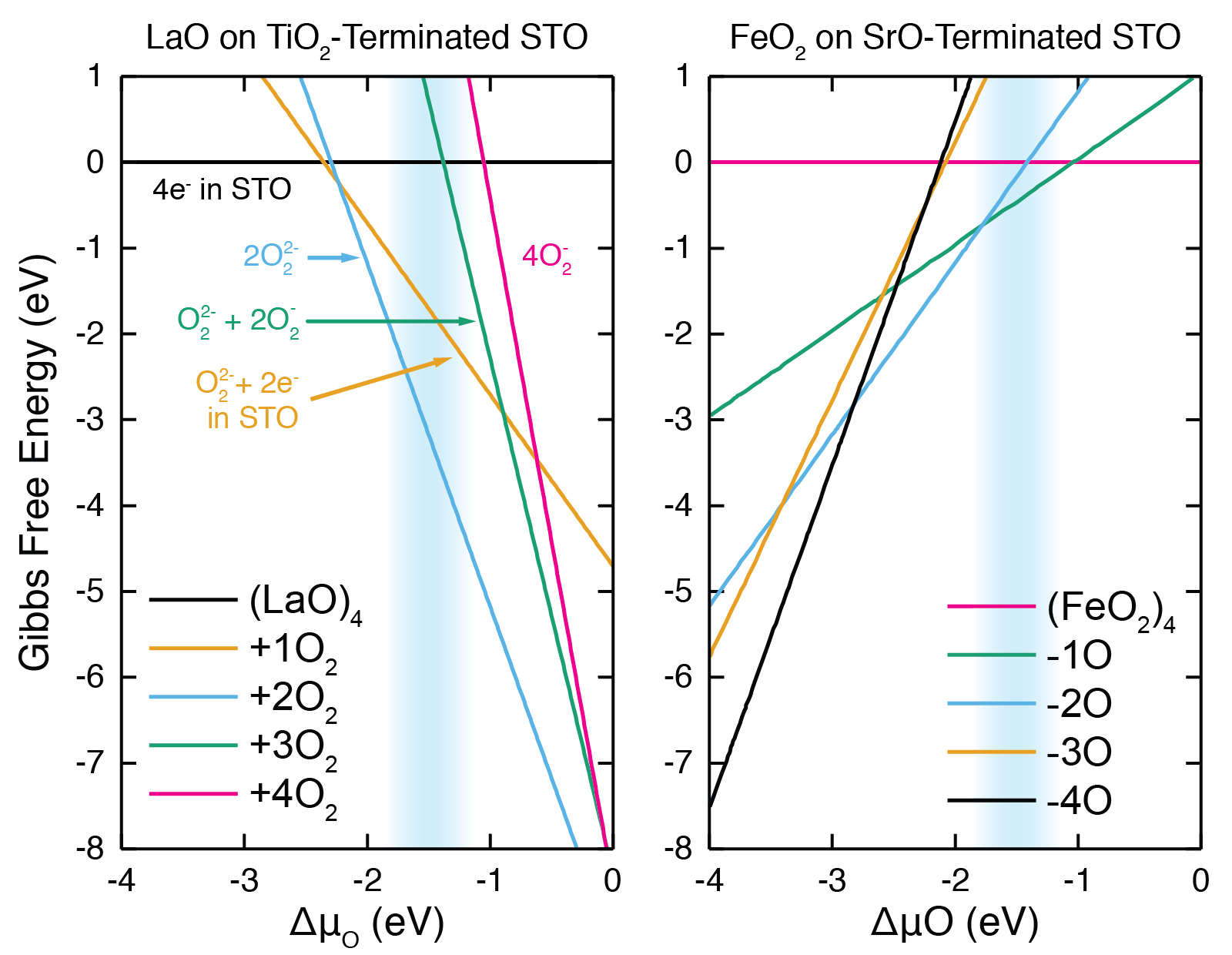}
\caption{Calculated Gibbs free energies for the LaO and FeO$_2$ planes (2$\times$2 lateral cell) deposited on the TiO$_2$-- (left) and SrO--terminated STO (right), respectively, as a function of oxygen chemical potential. Under our deposition conditions (indicated by the shaded region), the LaO film adopts an oxygen-rich configuration, while the FeO$_2$ plane is oxygen-deficient. \label{dft_gibbs}}
\end{figure}

Since La atoms have three valence electrons, the formation of the LaO film on TiO$_2$--terminated STO leaves one valence electron per La. According to our calculations, this remaining valence electron is either transferred to the STO conduction band or, in the presence of oxygen, gets trapped by the oxygen molecules that adsorb on the LaO film in the form of negatively charged oxygen species; the calculated Gibbs free energies for the latter mechanism are shown on the left side of Figure \ref{dft_gibbs}. In either case, the LaO plane as such is ordered and stoichiometric.

In contrast, the formation of an ordered stoichiometric FeO$_2$ plane on SrO--terminated STO is unlikely. Such a plane would feature Fe$^{4+}$ ions that are unstable in the bulk form, even under normal conditions.\cite{Falcon2002} In our case, deposition conditions result in the Fe$_2$O$_3$ composition of the plane, which corresponds to two oxygen vacancies per 2$\times$2 lateral cell (see right side of Figure \ref{dft_gibbs}). While an investigation of the dynamics of such a nominally FeO$_2$ / SrO interface is beyond the scope of the present work, it is clear that oxygen vacancies in this plane can facilitate complex rearrangements of oxygen atoms and cations that could amount to conversion of the FeO$_2$ / SrO interface to another structure. Similar trends are obtained using the \textsc{PBEsol}+$U$, as shown in the supporting information. In this case, the LaO + extra-O configurations become more stable because the +$U$(Ti) correction shifts the STO conduction band to higher energies, thus promoting electron transfer to the adsorbed oxygen species. Similarly, FeO$_2$ configurations with oxygen vacancies become more stable because +$U$(Fe) shifts occupied Fe 3$d$ states into the valence band, thus promoting charge transfer of the form Fe$^{4+}$ + O$^{2-}$ $\rightarrow$ Fe$^{3+}$ + O$^{-}$. This, in turn, promotes the formation of oxygen vacancies.

Our results illustrate the dramatic effects and non-equilibrium nature of shuttered growth, as well as the importance of thermodynamic and kinetic considerations to design targeted oxide heterostructures. STEM-EELS shows that shuttered MBE growth is able to produce exceptionally high-quality and defect-free LFO / $n$-STO interfaces. However, while AR-XPS indicates that two different STO substrate terminations were achieved prior to LFO deposition, atomic-scale composition mapping of the final as-grown heterojunctions reveals a LaO / TiO$_2$ interface structure for both cases. We observe no long-range film-substrate cation intermixing and minimal valence changes, suggesting that other factors must lead to the observed structure. \textit{Ab initio} simulations of interface stability show that FeO$_2$ / SrO is much less energetically preferred than LaO / TiO$_2$. We propose that the sequential nature of the shuttered growth mode may lead to unstable Fe$^{4+}$ ions that can drive the system toward a dynamic structural rearrangement \textit{via} oxygen vacancies. Further modeling of potential kinetic pathways and experimental study into the effects of shuttering sequence may open new ways to deterministically control the structure and properties of oxide interfaces.

\clearpage

\section*{Supporting Information}

Supporting information describing the methods used, as well as additional details about the EELS and DFT calculations is available.

\section*{Author Contributions}

S.R.S conducted STEM-EELS sample preparation, imaging, and analysis. P.V.S. conducted DFT calculations. R.B.C. synthesized the films. All authors originated the study, contributed to the data interpretation, discussions, and writing.

\section*{Competing Interest Statement}

The authors declare no competing financial interests.

\section*{Acknowledgements}

This work was supported by the U.S. Department of Energy, Office of Science, Division of Materials Sciences and Engineering under award \#10122. Growth was supported by the Linus Pauling Distinguished Postdoctoral Fellowship at Pacific Northwest National Laboratory (PNNL LDRD PN13100/2581). All work was performed in the Environmental Molecular Sciences Laboratory, a national science user facility sponsored by the Department of Energy's Office of Biological and Environmental Research and located at Pacific Northwest National Laboratory. Pacific Northwest National Laboratory (PNNL) is a multiprogram national laboratory operated for DOE by Battelle.

\clearpage

\bibliography{References}

\end{document}


\title{Dynamic Interface Rearrangement in LaFeO$_3$ / $n$-SrTiO$_3$ Heterojunctions}

\author{Steven R. Spurgeon}
\affiliation{Physical and Computational Sciences Directorate, Pacific Northwest National Laboratory, Richland, Washington 99352, USA}

\author{Peter V. Sushko}
\affiliation{Physical and Computational Sciences Directorate, Pacific Northwest National Laboratory, Richland, Washington 99352, USA}

\author{Ryan B. Comes}
\affiliation{Physical and Computational Sciences Directorate, Pacific Northwest National Laboratory, Richland, Washington 99352, USA}
\affiliation{Department of Physics, Auburn University, Auburn, Alabama 36849, USA}

\author{Scott A. Chambers}
\affiliation{Physical and Computational Sciences Directorate, Pacific Northwest National Laboratory, Richland, Washington 99352, USA}

\date{\today}

\maketitle

\section*{Supporting Information}

\subsection{Methods}

We have prepared LFO / $n$-STO (001) heterojunctions using oxygen-assisted MBE, as described elsewhere.\cite{Comes2016a} Several 0.05\% Nb-doped STO substrates were chemically treated to achieve a TiO$_2$ substrate termination, which we confirmed using AR-XPS. A single SrO layer was then deposited on a subset of the samples and the resulting termination was also confirmed \textit{via} AR-XPS. Finally, LFO was deposited in a shuttered growth sequence with the FeO$_2$ (LaO) layer deposited on SrO (TiO$_2$), respectively. Here we discuss the 9 unit cell (u.c.)-thick films; we call the FeO$_2$ / SrO configuration the ``$A$-Terminated'' sample and the LaO / TiO$_2$ configuration the ``$B$-Terminated'' sample. Details of the synthesis and characterization procedures are given in reference \citenum{Comes2016a}.

Cross-sectional STEM samples were prepared using an FEI Helios NanoLab Dual-Beam Focused Ion Beam (FIB) microscope and a standard lift out procedure along the STO [100] zone-axis, with initial cuts made at 30 kV / 2$^{\circ}$ and final polishing at 1 kV / 3$^{\circ}$ incidence angles. STEM-HAADF images and STEM-EELS maps were collected on a JEOL ARM-200CF microscope operating at 200 kV, with a convergence semi-angle of 27.5 mrad and an STEM-EELS collection angle of 42.9 mrad. STEM-EELS maps were collected using a $\sim1$ \AA \, probe size with a $\sim130$ pA probe current and a 0.25 eV ch$^{-1}$ dispersion, yielding an effective energy resolution of 0.75 eV. The composition maps shown in the supplemental section were acquired with a 1 eV ch$^{-1}$ dispersion and a 4$\times$ energy binning. No plural scattering correction was performed, since zero loss measurements confirm that the samples are sufficiently thin ($t/\lambda < 0.76$ inelastic mean free paths). The resulting spectrum images were processed to remove X-ray spikes and principal component analysis (PCA) was used to enhance the signal-to-noise of the composition maps in Figure 1. Figure 2 shows the raw, power-law background spectra extracted from each unit cell of the map.

LaFeO$_3$ / SrTiO$_3$ interfaces were represented using periodic slab model, where the LaFeO$_3$ and SrTiO$_3$ parts were each 4 u.c. thick. The 2$\times$2 lateral cell with the in-plane lattice parameter corresponding to bulk STO ($a$=$b$=3.905 \AA) was used. The supercell parameter along the $c$ axis was 50 \AA, which leaves a vacuum gap of over 20 \AA. The total energy of the system was minimized with respect to all degrees of freedom of the slab, unless stated otherwise. The calculations were performed using the Vienna \textit{Ab Initio} Simulation Package.\cite{Kresse1996,Kresse1999} The projected augmented wave method was used to approximate the electron-ion potential.\cite{Blochl1994} Exchange-correlation effects were treated with in the Perdew-Burke-Ernzerhoff (PBE) functional form, modified for solids.\cite{Perdew2008} The plane-wave basis with a 500 eV cutoff and the 2$\times$2$\times$1 Monkhorst-Pack $k$-point mesh were used. The charge and spin density distribution was analyzed using the Bader method.\cite{Bader1990,Henkelman_bader2009} The energies of self-consistent calculations were converged to 10$^{-5}$ eV/cell and the convergence of the total energy with respect to atomic coordinates was 10$^{-4}$ eV.

\subsection{Sr Composition Mapping}

\begin{figure}[h!]
\includegraphics[width=\textwidth]{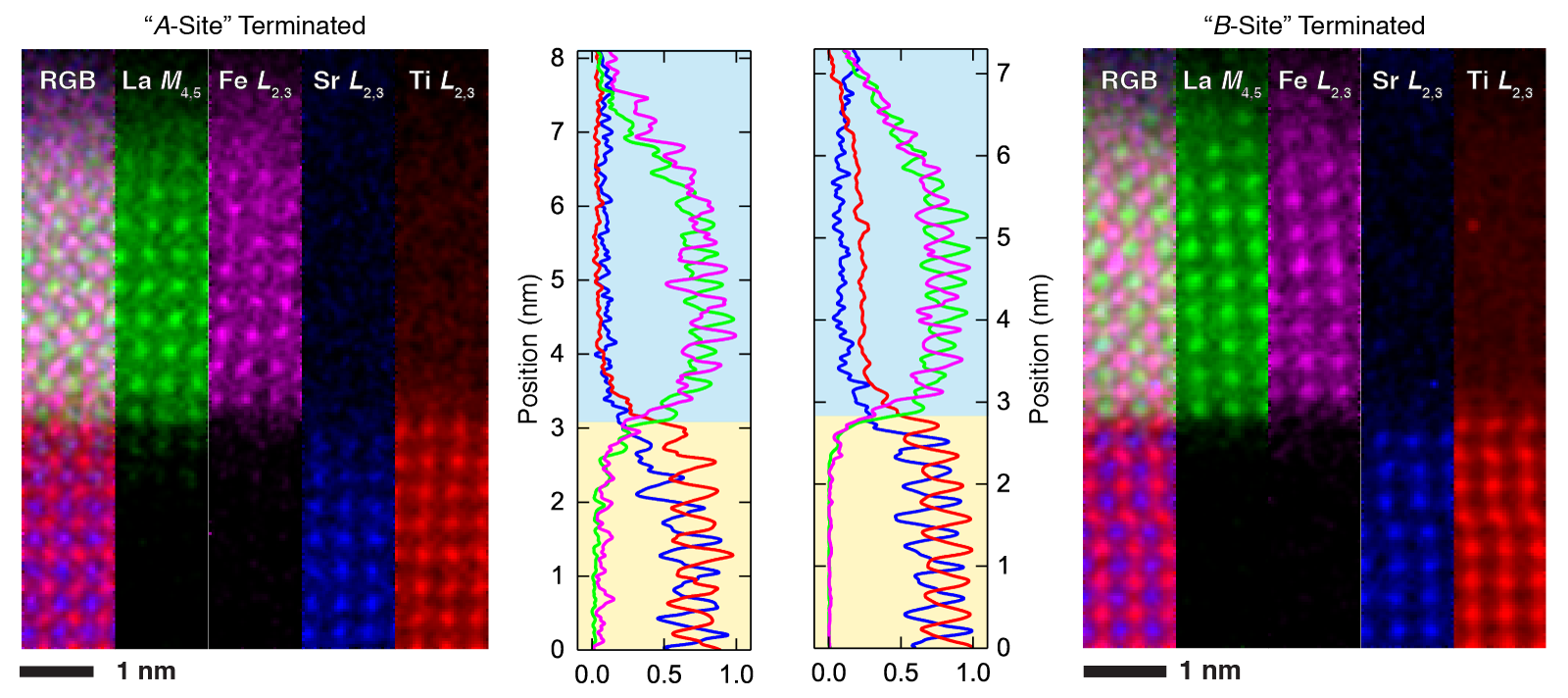}
\caption{Cross-sectional STEM-EELS mapping of local composition. Smoothed composite, La $M_{4,5}$, Sr $L_{2,3}$, Fe $L_{2,3}$, and Ti $L_{2,3}$ integrated signal maps, alongside the unprocessed integrated line profiles for the nominal ``$A$-site'' (left) and ``$B$-site'' (right) samples, respectively. The data have not been denoised in any way and the shaded regions mark the film-substrate interface.\label{eels_sr}}
\end{figure}

In addition to the STEM-EELS maps shown in the main text, we have conducted composition mapping at low dispersion to include all four alloy components. As shown in Figure \ref{eels_sr}, the same effective FeO$_2$ / LaO / TiO$_2$ / SrO interface stacking sequence is observed for both samples. We observe minimal Sr intermixing into the LFO film and intermixing trends for the other species that are comparable to those presented in the main text.

\newpage

\subsection{Defects in the LaO and FeO$_2$ planes on SrTiO$_3$ (001)}

It is well known that DFT functionals based on the generalized gradient approximation (GGA), including \textsc{PBEsol} functional used in this work, underestimate one-electron band gaps of insulators and semiconductors and lack accuracy in predicting spatial localization of electrons and holes. Incorrect positions of the band edges translate into erroneous relative energies of configurations where electrons occupy states of the conduction band (as in the case of LaO plane on TiO$_2$-terminated STO) and configurations where these electrons are trapped by the oxygen species adsorbed on the LaO plane. Similarly, in the case of an FeO$_2$ plane on the SrO-terminated STO, the relative energies of the stoichiometric plane that has formally Fe$^{4+}$ species and configurations of oxygen-deficient plane that has Fe$^{3+}$ are both affected. These deficiencies of GGA DFT can be mitigated to some extent using GGA+$U$ approach. 

\begin{figure}[h!]
\includegraphics[width=0.8\textwidth]{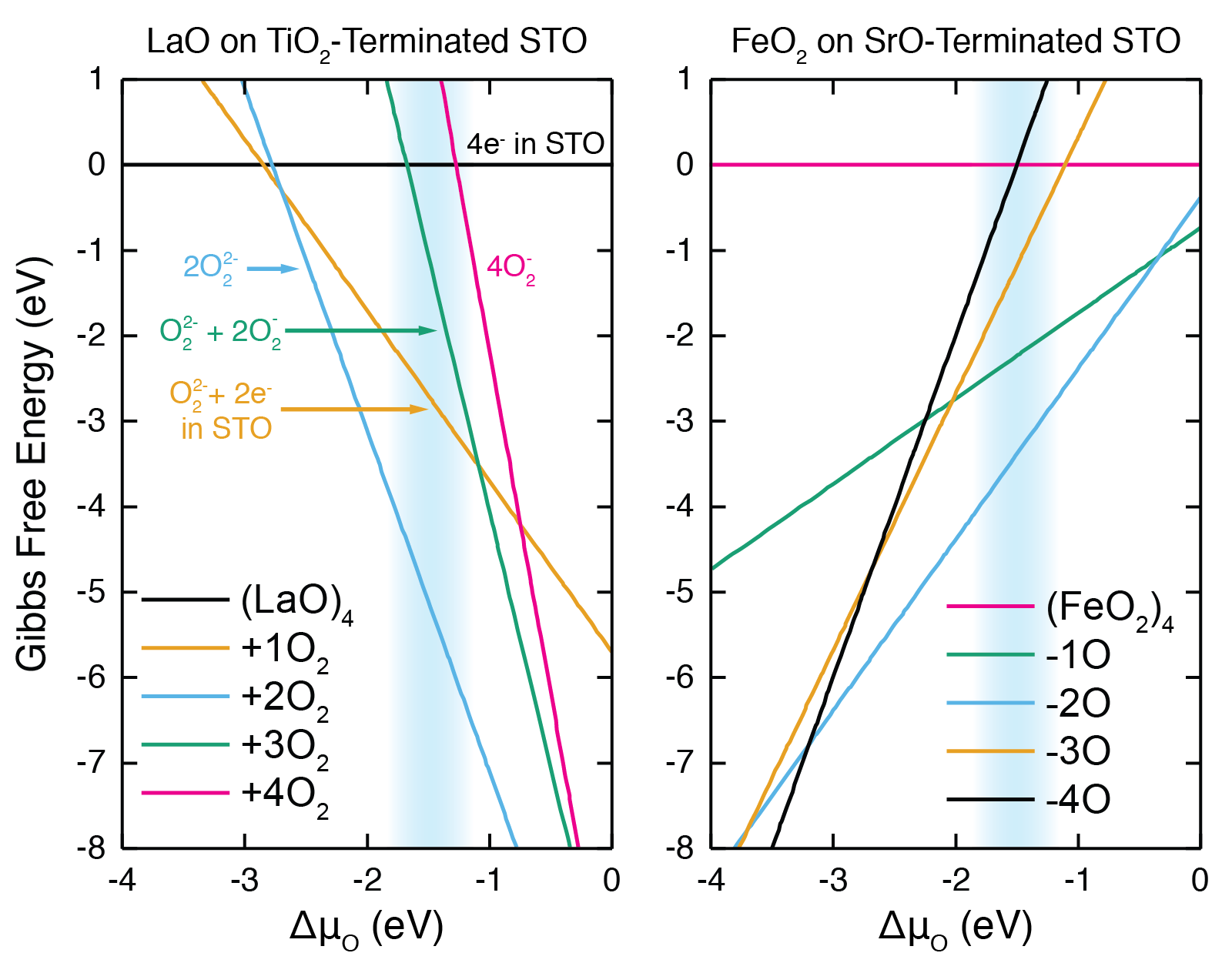}
\caption{Gibbs free energies calculated using \textsc{PBEsol}+$U$ for the LaO and FeO$_2$ planes deposited on the TiO$_2$--terminated STO (left) and SrO--terminated STO (right), respectively, as a function of oxygen chemical potential. The 2$\times$2 lateral cell was used in these calculations. Under our deposition conditions (indicated by the shaded region), the LaO film adopts an oxygen-rich configuration, while the FeO$_2$ plane is oxygen-deficient. \label{dft_gibbs_plusu}}
\end{figure}

Here we have investigated the effect of $U$ correction on the stability of oxygen-rich LaO / TiO$_2$ and oxygen-poor FeO$_2$ / SrO interfaces with respect to the corresponding stoichiometric systems. The $U$ correction was applied to the Ti and Fe 3$d$ atomic orbitals, as described in reference \citenum{Dudarev1998}. The calculated Gibbs free energies for the $U$=4 eV (see Figure \ref{dft_gibbs_plusu}) for the geometrical structures pre-optimized using the \textsc{PBEsol} functional. Comparison of these free energies with the data in Figure 4 of the main manuscript shows that trapping molecular oxygen at the LaO / TiO$_2$ interface and forming oxygen vacancies at FeO$_2$ / SrO interface is more favorable for positive $U$. While a determination of the optimal $U$ value is beyond the scope of the present work, these results support our conclusion that these two interfaces exhibit qualitatively different structures during shuttered growth mode deposition. This behavior, in turn, can have different effects on the substrate underneath.

\newpage

\begin{figure}[h!]
\includegraphics[width=\textwidth]{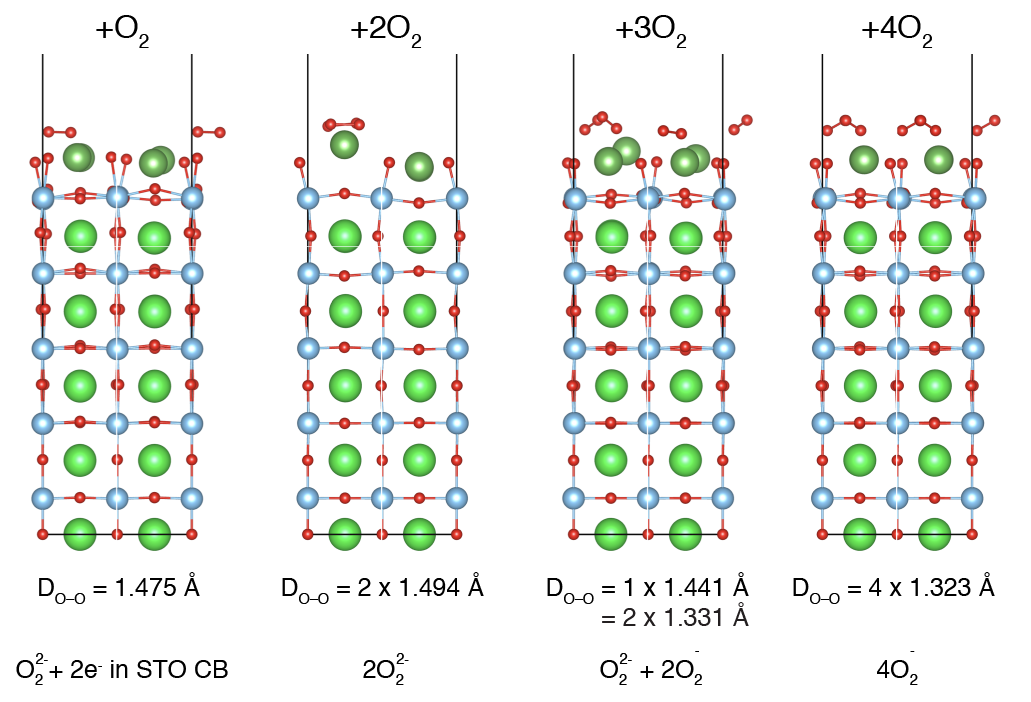}
\caption{Calculated configurations of the LaO plane with extra oxygen during the initial formation of the LFO film, corresponding to the cases given in main text Figure 4. Distances between oxygen atoms of the molecular ions and their formal ionic charges are shown at the bottom of the figure. Light green = Sr, dark green = La, blue = Ti, and red = O atoms.  \label{dft_configs}}
\end{figure}

\clearpage
\bibliography{References}